# Composition dependence of bulk properties in the Co-intercalated transition-metal dichalcogenide Co$_{1/3}$TaS$_2$

Pyeongjae Park[1,2,3*], Woonghee Cho[1,2], Chaebin Kim[1,2], Yeochan An[1,2], Maxim Avdeev[4,5], Kazuki Iida[6], Ryoichi Kajimoto[7], and Je-Geun Park[1,2,8*]

[1]*Center for Quantum Materials, Seoul National University, Seoul 08826, Republic of Korea*
[3]*Department of Physics and Astronomy, Seoul National University, Seoul 08826, Republic of Korea*
[3]*Materials Science and Technology Division, Oak Ridge National Laboratory, Oak Ridge, Tennessee 37831, USA*
[4]*Australian Nuclear Science and Technology Organization, Locked Bag 2001, Kirrawee DC, NSW 2232, Australia*
[5]*School of Chemistry, The University of Sydney, Sydney, NSW, 2006, Australia*
[6]*Comprehensive Research Organization for Science and Society (CROSS), Tokai, Ibaraki 319-1106, Japan*
[7]*Materials and Life Science Division, J-PARC Center, Tokai, Ibaraki 319-1195, Japan*
[8]*Institute of Applied Physics, Seoul National University, Seoul 08826, Republic of Korea*

*Corresponding Authors: parkp@ornl.gov & jgpark10@snu.ac.kr

## Abstract

Spontaneous Hall conductivity has recently been reported in the triangular lattice antiferromagnet Co$_{1/3}$TaS$_2$ under a zero magnetic field. This phenomenon originates from the distinctive noncoplanar triple-**Q** magnetic ground state, possessing uniform real-space Berry curvature characterized by scalar spin chirality. We investigated the physical properties of Co$_{1/3}$TaS$_2$ by judiciously controlling the composition, revealing a drastic change in its bulk properties, even by slight variations in cobalt composition, despite the same crystal structure. For $0.299 \leq x \leq 0.325$, Co$_x$TaS$_2$ keeps all the characteristics of the ground state consistent with the previous studies – two antiferromagnetic phase transitions at $T_{N1}$ and $T_{N2}$ ($< T_{N1}$), a large spontaneous Hall conductivity ($\sigma_{xy}(\mathbf{H} = 0)$), and a weak ferromagnetic moment along the c-axis. However, samples with $x \geq 0.330$ exhibit distinct bulk properties, including the absence of both $\sigma_{xy}(\mathbf{H} = 0)$) and the weak ferromagnetic moment. Our neutron diffraction data reveal that Co$_x$TaS$_2$ with $x \geq 0.330$ develops coplanar helical magnetic order with $\mathbf{q}_{m1}$ = (1/3, 0, 0). This is entirely different from what has been seen in $x \leq 0.325$, explaining the observed composition dependence.

With the advent of van der Waals magnets [1,2], there is a growing interest in finding new layered magnetic materials. Up until now, most of the research has been focused on a few selected classes of materials like TMPS$_3$ (TM = 3d transition metal) [3,4], CrI$_3$ [5], and Fe$_3$GeTe$_2$ [6,7], to name only a few. However, another new class of materials has recently gained attention as a layered magnet that can be exfoliated mechanically: intercalated transition-metal dichalcogenide (TMDC). TM$_{1/3}$(Ta,Nb)S$_2$ is one of the intercalated TMDC classes wherein the TM sites form a magnetic triangular lattice. Due to the geometrical frustration of the triangular lattice and the myriad combinations of different elements possible, TM$_{1/3}$(Ta,Nb)S$_2$ has provided a plethora of intricate phenomena stemming from various magnetic structures: a chiral soliton lattice in Cr$_{1/3}$NbS$_2$ [8-10], exchange bias induced by spin-glass in Fe$_{1/3}$NbS$_2$ [11], and spontaneous Hall conductivity ($\sigma_{xy}(\mathbf{H} = 0)$) despite vanishing net magnetization in Co$_{1/3}$(Ta,Nb)S$_2$ [12,13], to name only a few.

Nevertheless, owing to the nature of intercalation, defects are inevitable to some extent in TM$_{1/3}$(Ta,Nb)S$_2$. The most commonly found case is an imperfect composition of TM. While $x$ = 1/3 in TM$_x$(Ta,Nb)S$_2$ corresponds to a full occupancy of the 2$c$ Wyckoff sites and forms a stoichiometric structure, synthesized TM$_x$(Ta,Nb)S$_2$ crystals in reality do not precisely possess $x$ = 1/3. This means the presence of either vacancy at the 2$c$ sites (under-doped, $x$ < 1/3) or excess TM atoms intercalated into other sites (over-doped, $x$ > 1/3). Notably, such defects can give rise to nontrivial effects or novel magnetic ground states in frustrated magnets [14-20].

Despite extensive studies on TM$_{1/3}$(Ta,Nb)S$_2$ since the 1980s [21,22], the potential significance of the TM's composition only began to receive attention recently. Remarkably, the magnetic properties of TM$_{1/3}$(Ta,Nb)S$_2$ exhibit significant differences with even slight changes in the TM's composition. For example, in Fe$_{1/3}$NbS$_2$, a different magnetic ground state emerges depending on whether it is slightly under-doped ($x$ ~ 0.32) or over-doped ($x$ ~ 0.35) [23], leading to disparate physical properties [24]. Similarly, the transport characteristics of Co$_x$NbS$_2$ strongly depend on $x$ despite its proximity to 1/3 [25]. This exceptional sensitivity to composition warrants in-depth investigation in its own right but also offers an excellent avenue for engineering intriguing magnetic properties in TM$_{1/3}$(Ta,Nb)S$_2$. In particular, the latter can be achieved without introducing too many vacancies ($x$ < 1/3) or surplus TM atoms ($x$ > 1/3), thereby maintaining the system's closeness to ideal stoichiometry. For instance, the precise control of the Fe composition has enabled sophisticated manipulation of the electrical switching effect in Fe$_x$NbS$_2$ that holds potential for device applications [24,26].

Co$_{1/3}$TaS$_2$ – a member of the TM$_{1/3}$(Ta,Nb)S$_2$ family – has recently gained recognition as a host of a unique chiral magnetic ground state: the tetrahedral triple-$\mathbf{Q}$ magnetic ordering [27,28]. This structure represents one of the three fundamental spin configurations of a triangular lattice antiferromagnet [see Fig. 1 of Ref. [27]], which had not been previously identified in bulk materials until its recent observation in $Co_{1/3}TaS_2$. This triple-**Q** magnetic order generates a significant real-space Berry curvature characterized by its scalar spin-chirality, believed to be the source of the substantial $\sigma_{xy}(\mathbf{H}=0) \sim 70\ \Omega^{-1}\text{cm}^{-1}$ observed in $Co_{1/3}TaS_2$ (*i.e.,* topological Hall effect). Importantly, this triple-**Q** ordering emerges without the need for a magnetic field and even maintains its resilience against it [27], in contrast to triple-**Q** structures observed in other studies [29-31]. While the stabilization mechanism and phenomenological spin Hamiltonian for this ordering have been proposed based on the electronic structure of $Co_{1/3}TaS_2$ [27], the investigation of this magnetic structure is still in its nascent stages.

This letter presents comprehensive bulk characterizations and neutron diffraction measurements for $Co_xTaS_2$ with $0.299 < x < 0.34$, revealing substantial composition-dependent behavior. For $0.299 < x < 0.325$, a large $\sigma_{xy}(\mathbf{H}=0) \sim 70\ \Omega^{-1}\text{cm}^{-1}$ and a weak ferromagnetic moment ($M_z(\mathbf{H}=0) \sim 0.01\mu_B$) emerge with magnetic ordering characterized by the ordering wave vector $\mathbf{q}_{m2} = (1/2, 0, 0)$, consistent with the recent reports [13,27,28]. Conversely, a different magnetic ground state with $\mathbf{q}_{m1} = (1/3, 0, 0)$ is established for $0.330 < x < 0.34$, leading to considerably different bulk properties from the under-doped sample. Notably, this phase does not exhibit finite $\sigma_{xy}(\mathbf{H}=0)$ and $M_z(\mathbf{H}=0)$, thereby demonstrating a direct link between the magnetic structure of $\mathbf{q}_{m2} = (1/2, 0, 0)$ and these two quantities in $Co_{1/3}TaS_2$.

Polycrystalline and single-crystal $Co_xTaS_2$ were synthesized as described in Refs. [13,27] with details given in Supplementary Material [32]. Magnetic properties were measured using MPMS-XL5 and PPMS-14 with the VSM option (Quantum Design USA). Longitudinal and Hall resistivity of $Co_xTaS_2$ were measured using PPMS-9, PPMS-14, CFMS-9T (Cryogenic Ltd, UK), and a home-built cryostat. The measured Hall voltage was anti-symmetrized to get rid of longitudinal components. Spontaneous magnetic moment and Hall conductivity were measured after field-cooling the sample under 5 T. Heat capacity was measured using PPMS-14. A nonmagnetic contribution to the heat capacity was obtained from a combination of the Debye model and an electron's contribution linear with the temperature. We used two Debye temperatures ($\theta_{D1}= 270$ K and $\theta_{D2}= 611.7$ K) to account for the coexistence of heavy (Ta) and light (S) elements.

Powder neutron diffraction was conducted at the WOMBAT high-intensity powder diffractometer ($x = 0.330(4)$ and $0.340(4)$, $\lambda = 2.41$ Å) and the ECHIDNA high-resolution powder diffractometer ($x = 0.319(3)$, $\lambda = 2.4395$ Å) at ANSTO, Australia. For $x = 0.340(4)$ and $0.319(3)$, 20 g of powder was used, and 10 g for $x = 0.330(4)$. Rietveld refinements were done using Fullprof [33]; see Supplementary Material [32]. Single-crystal neutron diffraction data of $Co_xTaS_2$ ($x = 0.319(3)$ and $0.340(4)$) were obtained from the 4SEASONS time-of-flight spectrometer (J-PARC, Japan) [34]. The crystals were aligned so that the (*H*0*L*) and (*HHL*) planes were horizontal for $x$ = 0.319(3) and 0.340(4), respectively. For $x$ = 0.319(3), the data were collected with the incident neutron energy $E_i$ = 14.0 meV and the Fermi chopper frequency $f$ = 150 Hz. For $x$ = 0.340(4), we used $E_i$ = 14.98 meV and $f$ = 150 Hz. We used the Utsusemi [35] and Horace [36] for data analysis. All powder/single-crystal samples for the neutron scattering measurements were carefully inspected by measuring temperature-dependent magnetic susceptibility; see Fig. S1 [32].

For a systematic study of the composition dependence, we first verified the reliability and homogeneity of $x$ for the synthesized single crystals via several analyses (e.g., Fig. 1(b)); see Supplementary Materials [32]. We then investigated possible changes in the established crystal structure ($P6_322$) [13] or unexpected Co disorder depending on the Co composition across 0.299 < $x$ < 0.34. In particular, we carefully inspected the intercalation profile of Co atoms by assessing the sharpness of the (101) Bragg peak and the Raman peak at 137 cm$^{-1}$. These peaks correspond to the superlattice peak from Co intercalation and the phonon mode predominantly governed by in-plane vibrations of Co atoms [37]. As shown in Fig. 1(c)–(d), both the measured powder XRD and Raman spectra remained consistent across different $x$ values, indicating a uniform crystal structure without noticeable disorder for 0.31 < $x$ < 0.34. This observation is corroborated by the near-resolution-limit full widths at half maximum (FWHMs) of the (101) XRD peak and 137 cm$^{-1}$ Raman peaks (Fig. 1(e)). On the other hand, a slight increase in FWHMs is discernible in $x$ < 0.31, which could be attributed to lattice inhomogeneity stemming from increased Co vacancies.

However, transition temperatures and detailed bulk properties vary significantly with $x$, even though the investigated $x$ range does not deviate much from the ideal value of $x$ = 1/3. Fig. 2 shows the temperature-dependent magnetization and resistivity of Co$_x$TaS$_2$ with 0.299 < $x$ < 0.34. Co$_{0.330(4)}$TaS$_2$, which is closest to the ideal composition, exhibits a single antiferromagnetic phase transition at 35 K [Fig. 2(d) and 2(j)]. When $x$ decreases below 0.330, the phase transition at 35 K splits into two at $T_{N1}$ and $T_{N2}$ ($T_{N2}$ < $T_{N1}$), as evident in Fig. 2(a)–(c) and 2(h)–(i). This particular composition range shows the same bulk properties as the recent studies [13,27,28]: a substantial $\sigma_{xy}(\mathbf{H} = 0)$ and a weak ferromagnetic moment below $T_{N2}$. While $T_{N1}$ = 38 K and $T_{N2}$ = 26.5 K remain nearly constant for 0.31 < $x$ < 0.325, $T_{N2}$ exhibits a decrease for $x$ < 0.31 [Fig. 2(a)], consistent with the proposed increase in the Co vacancy effect on the XRD and Raman spectra for $x$ < 0.31 (Fig. 1(e).

For the over-doped samples ($x$ > 1/3), $T_N$ gets significantly increased by a slight Co composition change. For instance, it increases by approximately 15 % (35 K → 41 K) for a 3 % increase of $x$ from 0.330(4) to 0.340(4). Hence, both under-doping and over-doping away from $x$ = 1/3 increase $T_N$ of $Co_{1/3}TaS_2$. This observation suggests an unusual composition dependence, as defects are generally anticipated to decrease $T_N$ by disrupting the formation of long-range order [23]. In addition, a steep increase of magnetization along the c-axis arises at 41 ~ 43 K [Fig. 2(e)–(f)], slightly above the main phase transition at 35 ~ 41 K [Figs. 2(e)–(f) and 2(k)–(l)]. However, this characteristic is continuously suppressed [Fig. 2e] and ultimately vanishes near $x \sim 1/3$ [Fig. 2d]. Therefore, this might originate from interstitial Co, occupying positions other than 2c Wyckoff sites and thus could act like strong magnetic impurities [23].

The most striking composition dependence manifests in the absence of the weak ferromagnetic moment ($M_z(\mathbf{H} = 0)$) and spontaneous Hall resistivity ($\rho_{xy}(\mathbf{H} = 0)$) for $x > 0.325$. Fig. 3 illustrates this through representative examples: $x = 0.325(4)$ and $0.336(5)$ (see also Fig. S2). Unlike the case of $x = 0.325(4)$, both finite $M_z(\mathbf{H} = 0)$ and $\rho_{xy}(\mathbf{H} = 0)$ are absent in $x = 0.336(5)$ [Fig. 3(b)–(c)]. A thorough examination of $M_z(\mathbf{H} = 0)$ and $\rho_{xy}(\mathbf{H} = 0)$ across all compositions reveals their presence only for $x < 0.325$ while being completely absent for $x > 0.330$. These two distinct phases seem to compete at $x = 0.327(4)$ [Fig. 3(d)]. Although the sample with $x = 0.327(4)$ undergoes a phase transition at 28 K and exhibits nonzero $M_z(\mathbf{H} = 0)$, the magnitude of $M_z(\mathbf{H} = 0)$ is markedly suppressed and $\rho_{xy}(\mathbf{H} = 0)$ approaches zero.

The field dependence of $M_z$ and $\rho_{xy}$ for $x = 0.336(5)$ deviates from a linear behavior below $T_N = 35$ K near zero magnetic fields [Fig. 3(b)–(c)], leading to the spike-like profile in $M_z(T)$ in Fig. 2(e)–(f). Consequently, these spike-like signals should not be misinterpreted as nonzero $M_z(\mathbf{H} = 0)$ despite their visual similarity to $M_z(T)$ curves from under-doped samples [Fig. 2(a)-(b)] that really have finite $M_z(\mathbf{H} = 0)$ below $T_{N2}$. Moreover, the temperature dependence of normal Hall coefficients for $x = 0.325(4)$ and $x = 0.336(5)$ demonstrates a large difference from each other at low temperatures [Fig. 3(e)]. Notably, the coefficient sharply increases below $T_N$ only in the over-doped sample. A similar feature is also found in the heat capacity of $x = 0.336(5)$ [Fig. 3(f)], showing that the release of magnetic entropy is confined to a narrow interval around $T_N = 35$ K. Coupled with a sudden resistivity drop at $T_N$ [Fig. 2(k)-(l)], this behavior implies that the phase transition in over-doped $Co_xTaS_2$ may exhibit a strong first-order nature. A sharp heat capacity peak, albeit less pronounced, is also observed in $x = 0.325(4)$ at $T_{N2}$. This is consistent with the expectation that the single-$\mathbf{Q}$–to–triple-$\mathbf{Q}$ transition at $T_{N2}$ should be first-order due to their discontinuous symmetry relation. We note that the total magnetic entropies of both samples are nearly identical.

To better understand the relation between $x$ and the presence of $\sigma_{xy}(\mathbf{H} = 0)$ in $Co_xTaS_2$, we studied the magnetic structure of three samples with $x = 0.319(3)$, $0.330(4)$, and $0.340(4)$ using neutron diffraction. Note that only the powder diffraction data of $x = 0.319(3)$ [Fig. 4(c)] includes weak $\lambda/2$ harmonic signals of ~0.3 %, leading to additional spurious peaks in the diffraction pattern compared to those of $x = 0.340(4)$ and $0.330(4)$ [Fig. 4(a)–(b)]. The profiles of $x = 0.340(4)$ and $0.330(4)$ are not contaminated by such $\lambda/2$ signals due to the pyrolytic graphite filter installed at WOMBAT. These three samples exhibit the same powder diffraction patterns in their paramagnetic phases, reaffirming their uniform crystal structure (see Fig. S3 [32]). However, the magnetic reflections of these samples reveal distinct magnetic structures for over-doped ($x \geq 0.330$) and under-doped ($x \leq 0.325$) $Co_xTaS_2$. Each case is described by different ordering wave vectors: $\mathbf{q}_{m1} = (1/3, 0, 0)$ and $\mathbf{q}_{m2} = (1/2, 0, 0)$, as demonstrated in their single-crystal diffraction spectra [Figs. 4(d)-(e), see also Fig. S4 [32]]. Notably, neither case is consistent with $\mathbf{q}_m=(1/3, 1/3, 0)$ as previously reported [38], *i.e.*, the 120° magnetic structure. A detailed discussion of this discrepancy can be found in Ref. [27].

The magnetic structure of $\mathbf{q}_{m2}$ in under-doped $Co_xTaS_2$ ($x \leq 0.325$) corresponds to a collinear single-**Q** configuration for $T_{N2} < T < T_{N1}$ [Fig. 4(g)] and a tetrahedral triple-**Q** configuration for $T < T_{N2}$ [Fig. 4(h)], as already reported in Refs. [27] and [28]. For over-doped $Co_xTaS_2$ ($x \geq 0.330$), the Rietveld refinement (solid black lines in Fig. 4(a)–(b)) gives the helical ordering illustrated in Fig. 4(f) as the most likely magnetic structure (see Fig. S5). In this spin configuration, magnetic moments lie on the b*-c plane and form a 150° angle with their counterparts on the adjacent Co layer, consistent with antiferromagnetic inter-layer interactions in $Co_{1/3}TaS_2$ [27] (see Fig. S6). In addition, we examined whether the observed anomalous increase of magnetization in Fig. 2(e)–(f) (*i.e.* the spike-like signal) affects the magnetic structure by comparing the diffraction pattern of $x = 0.330(4)$ with $x = 0.340(4)$ [Fig. 4(a)–(b)]. As a result, no noticeable difference was found in their overall diffraction profiles, indicating identical magnetic structures. However, these two samples possess different magnitudes of an ordered moment: $1.37(3)\mu_B$ for $x = 0.330(4)$ and $1.79(4)\mu_B$ for $x = 0.340(4)$. Such a difference is significant compared to the slight difference in $x$, similar to the considerable increase of $T_N$ from $x = 0.330(4)$ to $0.340(4)$. It is unanswered how such a minor excess Co could lead to such enhancement. Another notable observation involves a large inconsistency in background levels below and above $T_N$ for $x = 0.330(4)$ and $0.340(4)$, detailed discussions of which can be found in Supplementary Material [32].

Fig. 4(i) and 4(j) show two phase diagrams summarizing our experimental findings, demonstrating the pronounced composition dependence of the bulk properties in $Co_xTaS_2$. Indeed, both magnetic structures with $\mathbf{q}_{m1}$ and $\mathbf{q}_{m2}$ arise from the $x$ value close enough to 1/3 and can be considered the intrinsic ground state of ideal $Co_{1/3}TaS_2$. The absence of $\sigma_{xy}(\mathbf{H} = 0)$ and $M_z(\mathbf{H} = 0)$ for the ground state with $\mathbf{q}_{m1}$ suggests that the magnetic structure with $\mathbf{q}_{m2}$ plays an important role in generating finite $\sigma_{xy}(\mathbf{H} =$

0) and $M_z(\mathbf{H} = 0)$. This is consistent with the recent studies proposing that the observed $\sigma_{xy}(\mathbf{H} = 0)$ originates from real-space Berry curvature within its noncoplanar triple-$\mathbf{Q}$ magnetic structure [see Fig. 4(h)] [27,28]. Conversely, the coplanar helical order described by $\mathbf{q}_{m1}$ lacks components generating real-space Berry curvature. Another noteworthy result is the reduction of $\sigma_{xy}(\mathbf{H} = 0)$ by lowering $x$ below ~0.31. Notably, this reduction coincides with the decrease of $T_{N2} = 26.5$ K [Fig. 4(i)]. This could be attributed to a non-negligible Co vacancy for $x < 0.31$, locally preventing the otherwise perfect formation of the all-in-all-out spin configuration depicted in Fig. 4(h). On the other hand, $M_z(\mathbf{H} = 0)$ remains nearly intact even for $x < 0.31$. Further interpretation of this observation is limited by the unknown nature of $M_z(\mathbf{H} = 0)$, which could be either an orbital magnetic moment from real-space Berry curvature or a spin moment from a slight out-of-plane canting [27].

Our work emphasizes two crucial points for $Co_{1/3}TaS_2$: i) careful attention must be given to the Co composition when investigating $Co_{1/3}TaS_2$, and ii) the Co composition can be an effective degree of freedom for tuning the magnetism of $Co_{1/3}TaS_2$ without deviating significantly from the ideal composition. The most straightforward explanation for the latter findings is that a slight tuning of the Fermi level ($E_f$) due to a different Co composition (intercalation of one Co atom provides two electrons to the system) has changed the momentum-dependent spin susceptibility at $E = E_f$ ($\chi(\mathbf{Q}, E_f)$), thereby leading to a different ordering wave vector of the magnetic structure. Indeed, delicate bulk properties in other isostructural compounds were also attempted to be described using the $E_f$ shift [25,39]. Notably, a recent theoretical study on $Co_{1/3}NbS_2$ – facing a situation similar to $Co_{1/3}TaS_2$ – demonstrated that hole doping (equivalent to removing Co) can induce $\mathbf{q}_{m2} = (1/2, 0, 0)$ [40], consistent with our observation of $\mathbf{q}_{m2} = (1/2, 0, 0)$ in under-doped $Co_{1/3}TaS_2$. However, the observed significant composition dependence appears to exceed what a mere $E_f$ shift would entail, considering a tiny change of $x$ from 0.330(4) to 0.325(4). This begs an alternative possibility beyond the simple $E_f$ shift: for whatever reasons, a slight change of Co composition might have considerably deformed the electronic structure near $E_f$. Interestingly, similar observations have been made in $Fe_{1/3}NbS_2$ very recently: the measured electronic structure suddenly changes when the composition crosses $x = 1/3$ [41]. Thus, examining the electronic structure of both under- and over-doped $Co_{1/3}TaS_2$ is imperative for a deeper understanding. This is further accentuated by the fact that the stability mechanism of the tetrahedral triple-$\mathbf{Q}$ ordering relies on nesting and is, therefore, heavily influenced by the Fermi surface geometry [27,42]. A different approach would involve interpreting our observation based on spin Hamiltonian. Indeed, the magnetic ground states of $\mathbf{q}_{m1} = (1/3, 0, 0)$ and $\mathbf{q}_{m2} = (1/2, 0, 0)$ are adjacent to each other in the phase diagram of a triangular lattice $J_1$-$J_2$-$J_3$ model [43]. Measuring the spin-wave spectra of both under- and over-doped samples using inelastic neutron scattering will be a promising future research for this approach. Finally, exploring the $x$ region below $x = 0.299$ represents another valuable avenue of investigation. This could offer additional insights into the composition-dependent magnetic order in $Co_{1/3}TaS_2$, particularly regarding the 120º magnetic order reported in $Co_{0.29}TaS_2$ decades ago [22,38].

**Acknowledgments**

We should acknowledge our numerous discussions with Cristian Batista, Yoon-Gu Kang, Myung Joon Han, Han-Jin Noh, and Ki Hoon Lee. This work was supported by the Samsung Science & Technology Foundation (Grant No. SSTF-BA2101-05). P. Park acknowledges support by the U.S. Department of Energy, Office of Science, Basic Energy Sciences, Materials Science and Engineering Division. The neutron scattering experiment at the Japan Proton Accelerator Research Complex (J-PARC) was performed under the user program (Proposal No. 2021B0049 and 2023A0036). J.-G.P. is partly funded by the Leading Researcher Program of the National Research Foundation of Korea (Grant no. 2020R1A3B2079375).

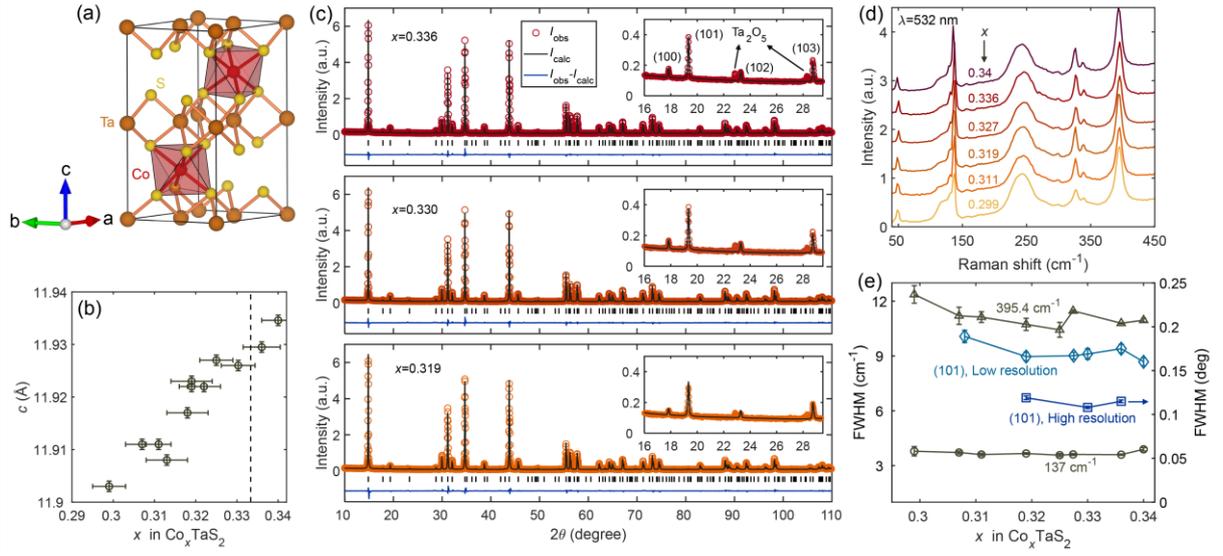

FIG. 1. Characterization of the Co$_x$TaS$_2$ structure (0.299 < $x$ < 0.34). (a) The crystallographic unit cell of Co$_{1/3}$TaS$_2$ where Co atoms are intercalated into the 2$c$ Wyckoff sites. (b) Comparison between the measured Co composition and the lattice parameter $c$ in single-crystal Co$_x$TaS$_2$. (c) High-resolution powder XRD data of three Co$_x$TaS$_2$ samples with $x$ = 0.336(5), 0.330(4), and 0.319(3). The Rietveld refinement results are shown as solid black lines. The inset illustrates identical (1, 0, $L$) superlattice peak profiles of the three samples with different $x$ values. (d) Raman spectra of single-crystal Co$_x$TaS$_2$ at 300 K. (e) Composition-dependent FWHMs of the (1, 0, 1) XRD peak and the Raman peaks at 137 and 395.4 cm$^{-1}$. Except for $x$ = 0.299, no significant increase in FWHMs is observed.

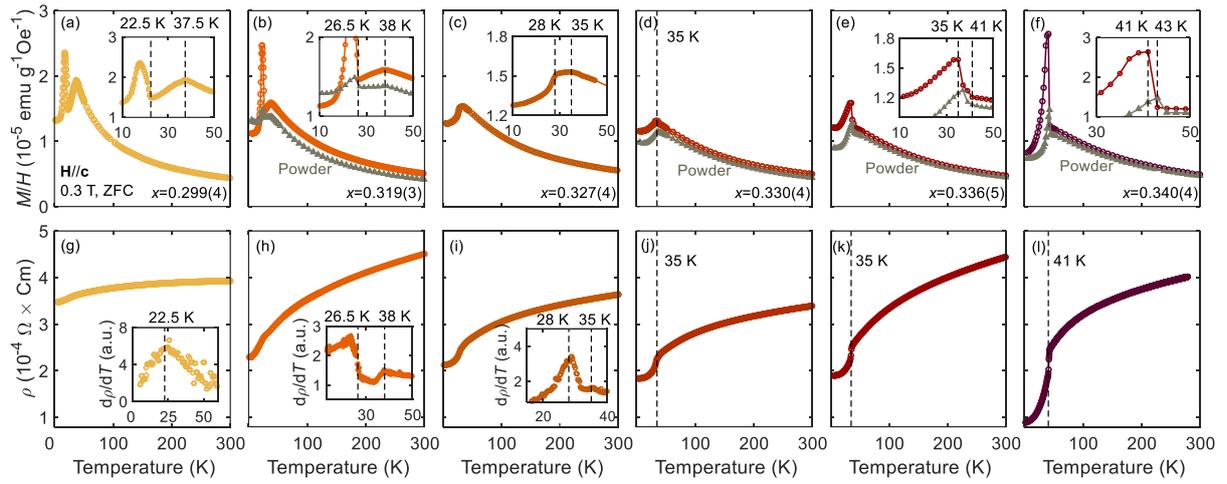

FIG. 2. Significant composition dependence of bulk properties in $Co_xTaS_2$. (a)–(f) The temperature-dependent magnetization of single-crystal $Co_xTaS_2$ along the c-axis, measured under a magnetic field of 0.3 T after zero-field cooling. Grey data points in each panel correspond to data from the powder sample with the same composition $x$. (g)–(l) Temperature-dependent resistivity ($\rho_{xx}$) of $Co_xTaS_2$. Insets focus on the temperature range surrounding the magnetic phase transitions, revealing a two-step transition process for specific composition values.

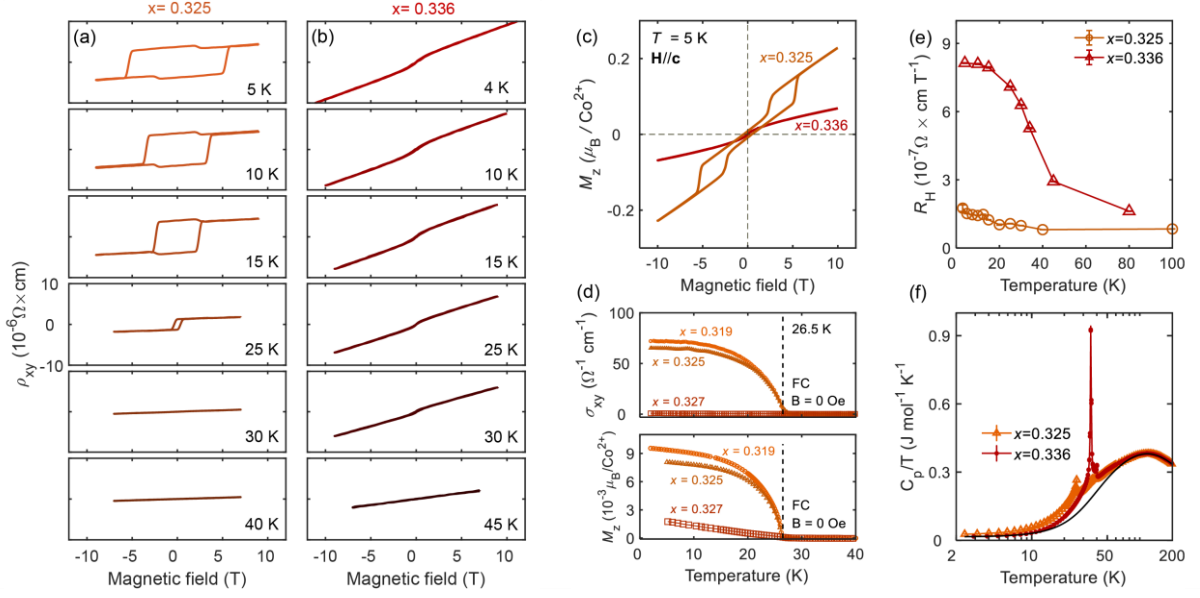

FIG. 3. Thorough comparison between under-doped ($x \leq 0.325$) and over-doped ($x \geq 0.330$) $Co_xTaS_2$. (a)–(c) The field dependence of the Hall resistivity ($\rho_{xy}(H)$) and the $c$-axis magnetization ($M_z(H)$) for $x = 0.325(4)$ and $0.336(5)$. (d) Temperature dependence of $M_z(H = 0)$ and $\sigma_{xy}(H = 0)$ in $Co_xTaS_2$ with $x \leq 0.327$. For $x \geq 0.327$, both quantities are measured as zero and thus are not shown here; see Figs. 4(i)–(j). (e)–(f) Temperature-dependent normal Hall coefficients and heat capacity of $Co_xTaS_2$ with $x = 0.325(4)$ and $0.336(5)$. The solid black line in (f) represents the nonmagnetic contribution of $C_p/T$.

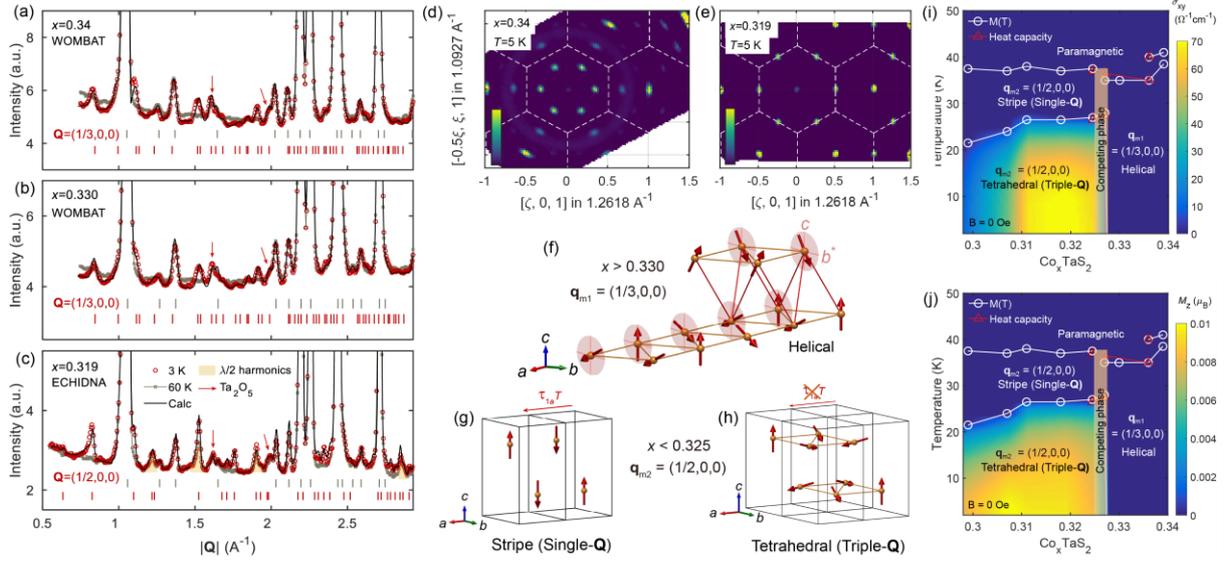

FIG. 4. Composition-dependent magnetic ground states in $Co_xTaS_2$. (a)–(c) Powder neutron diffraction patterns of $Co_xTaS_2$ with $x = 0.340(4)$, $0.330(4)$, and $0.319(3)$ above and below $T_N$. Some differences in the diffraction spectra between (a)–(b) and (c) result from the weak $\lambda/2$ signal only in (c); see Fig. S3 [32]. Solid black lines in (a)–(c) represent simulated nuclear and magnetic diffraction patterns with the spin configurations shown in (f) ($x = 0.340(5)$ and $0.330(4)$) and (h) ($x = 0.319(3)$). Grey (red) vertical ticks denote the positions of nuclear (magnetic) reflections. (d)-(e) Single-crystal neutron diffraction spectra of $Co_xTaS_2$ with $x = 0.340(4)$ and $x = 0.319(3)$ at 5 K. (f) The magnetic structure of $Co_xTaS_2$ with $x = 0.340(4)$ or $0.330(4)$ obtained from Rietveld refinement. (g)–(h) The magnetic structure of $Co_xTaS_2$ with $0.319(3)$ for $T_{N2} < T < T_{N1}$ and $T < T_{N2}$, respectively. (i)–(j) Phase diagram of $Co_xTaS_2$ over the entire composition. Colour plots in (i) and (j) display measured $M_z(\mathbf{H} = 0)$ and $\sigma_{xy}(\mathbf{H} = 0)$ after field cooling as described in the main text.